\documentclass{article}

\def\be{\begin{equation}}
\def\ee{\end{equation}}
\def\bea{\begin{eqnarray}}
\def\eea{\end{eqnarray}}

\def\d{\partial}
\def\R{{\bf R}}
\def\tr{{\rm tr}}
\def\fr#1#2{{\textstyle{#1\over#2}}}

\begin{document}

\title{Cosmological models and centre manifold theory}

\date{}

\author{Alan D. Rendall
\\ Max-Planck-Institut f\"ur Gravitationsphysik
\\ Am M\"uhlenberg 1, 14476 Golm, Germany
}

\maketitle

\begin{abstract}
Centre manifold theory is applied to some dynamical systems arising
from spatially homogeneous cosmological models. Detailed 
information is obtained concerning the late-time behaviour of
solutions of the Einstein equations of Bianchi type III with 
collisionless matter. In addition some statements in the literature
on solutions of the Einstein equations coupled to a massive
scalar field are proved rigorously.
\end{abstract}

\section{Introduction}\label{intro}

Over the years a great deal of effort has been put into the study of
homogeneous solutions of the Einstein equations coupled to various
matter sources. This class of solutions has the advantage that the 
evolution equations reduce to ordinary differential equations. This
allows a mathematically rigorous treatment of many issues for which
corresponding results seem out of reach at present when partial
differential equations are involved. The area of mathematics which
is relevant for the study of the time evolution of homogeneous 
cosmological models is the theory of dynamical systems. An excellent 
source of background information on the application of dynamical
systems to cosmology is the book edited by Wainwright and Ellis
\cite{wainwright97}. 

When a dynamical system (i.e. a system of ordinary differential
equations) constituting a model of an aspect of the real world
has been derived a natural task is to understand the qualitative
behaviour of general solutions of the equations. Doing this may
involve combining many techniques. In the following we concentrate
on one particular type of technique. In understanding a dynamical
system it is useful to determine the stationary solutions. Once 
these have been found it is desirable to know the behaviour of
solutions which remain in a neighbourhood of one of these stationary
solutions. If the system is linearized about the stationary point
then there is often a relationship between solutions
of the linearized system and solutions of the original system which
remain near the stationary point. In the case of hyperbolic stationary
points (defined below) this works very well. When a stationary point
is not hyperbolic things are not so easy. In the latter case an 
important role is played by centre manifold theory. This theory will
be illustrated by various cosmological examples in what follows.

One example of an application of centre manifold theory to homogeneous 
cosmological models in the literature can be found in \cite{nilsson00}. 
There it was used to determine the asymptotic behaviour of spacetimes of 
Bianchi type VII${}_0$ with a radiation fluid. A feature of this example 
which is typical is that it represents a degenerate or borderline case. 
The case of a perfect fluid with equation of state $p=(\gamma-1)\rho$ 
and $\gamma\ne 4/3$ can be analysed without centre manifold theory 
\cite{wainwright99} and it is found that the behaviour changes 
qualitatively when passing through the case of the radiation fluid 
($\gamma=4/3)$.

In this paper two different applications of centre manifold theory are
considered. The first is to homogeneous spacetimes with collisionless
matter. More specifically, it concerns locally rotationally symmetric (LRS)
models of Bianchi type III. These play the role of a degenerate case
within more general classes of Bianchi models with collisionless matter
studied in \cite{rendall99a} and \cite{rendall00c}. The centre manifold
analyses of the present paper extend and complete the results of 
\cite{rendall99a} and \cite{rendall00c}. The other application is to 
homogeneous and isotropic spacetimes with a massive scalar field. The 
centre manifold analysis provides a rigorous confirmation of heuristic 
conclusions on inflation in this class of spacetimes due to Belinskii et. al. 
\cite{belinskii86}.

The results on models with collisionless matter obtained in the following
address a question raised in \cite{rendall00c}. Suppose we have an
expanding cosmological model with collisionless matter described by the
Vlasov equation. The particles composing the matter have a certain
non-zero velocity dispersion. It might be expected that if these particles
are massive the velocity dispersion will decay with time as a 
consequence of the cosmological expansion so that the spacetime will
look more and more like a dust model in the phase of unlimited expansion.
It was shown in \cite{rendall00c} that this is indeed the case for
LRS models of Bianchi types I and II. For Bianchi type III, however,
despite the amount of information on the dynamics obtained in 
\cite{rendall00c}, this issue was not settled. The reason for the difficulty 
is that in Bianchi type III one of the scale factors has an anomalously slow 
rate of expansion. In fact the results of \cite{rendall00c} did not even 
suffice to show that this scale factor is unbounded. It is shown below that 
dust-like asymptotics also occurs in the case of Bianchi type III and that
the problematic scale factor does increase without limit. An 
analogous result for massless particles is also obtained, thus completing 
the results of \cite{rendall99a}. It is to be hoped that these results will 
also open the way to further progress in understanding inhomogeneous 
models with collisionless matter such as those discussed in 
\cite{andreasson01}.

The second section contains some general information about centre
manifold theory. Some equations for spacetimes of Bianchi type III
with kinetic matter are recalled in Section \ref{setupmassive}.
Centre manifold theory is applied to Bianchi III spacetimes
with collisionless matter in the case of massive particles in Section 
\ref{cmassive} and in the case of massless particles in Section 
\ref{cmassless}. Section \ref{scalar} contains results on homogeneous 
and isotropic solutions with a massive scalar field. The final section 
presents some insights which can be obtained from the results in the body 
of the paper.     

\section{Centre manifold theory}

Let $\dot x=f(x)$ be a system of ordinary differential equations, where 
$x$ is a point of $\R^n$ and the function $f$ is smooth ($C^\infty$). 
Consider a
point $x_0$ with $f(x_0)=0$. Then $x(t)=x_0$ is a time-independent
solution of the system. In other words, $x_0$ is a stationary point of
the given dynamical system. The linearization of the system about 
$x_0$ is the system $\dot y=Ay$ where the matrix $A$ has entries
$A_i^j=\d f^j/\d x^i$. If each eigenvalue $\lambda$ of $A$ has the
property that the real part of $\lambda$ is non-zero then the
stationary point is called hyperbolic. In that case the theorem
of Hartman and Grobman (\cite{perko01}, p. 120) shows that the linearized 
system is topologically equivalent to the original system in a neighbourhood
of $x_0$. In other words, there is a homeomorphism of a neighbourhood
of $x_0$ onto a neighbourhood of the origin in $\R^n$ such that solutions
of the non-linear system are mapped onto solutions of the linearized
system. Note that this homeomorphism cannot in general be improved to
a $C^2$ diffeomorphism. (See \cite{perko01}, p. 127.) 

When the matrix $A$ has eigenvalues which are purely imaginary and, in
particular, when it has zero as an eigenvalue things are more complicated
than for hyperbolic stationary points. Recall that the generalized 
eigenspace associated to an eigenvalue $\lambda$ of $A$ consists of all
complex vectors $x$ such that $(A-\lambda I)^k x=0$ for some positive
integer $k$. The centre subspace $E_c$ of the dynamical system at
$x_0$ is defined as follows. Take the direct sum of the generalized 
eigenspaces corresponding to all purely imaginary eigenvalues and then 
restrict to the real subspace. Then $E_c$ is an invariant subspace of
$A$ and the restriction of $A$ to $E_c$ has purely imaginary eigenvalues.
$E_c$ is the maximal subspace having this property. The centre manifold
theorem \cite{carr} says that there is a manifold $M_c$ passing through
$x_0$ which is invariant under the flow of differential equation and whose
tangent space at $x_0$ is $E_c$. The manifold $M_c$ is called a centre 
manifold. It can be chosen to be $C^k$ for any positive integer $k$ but not 
necessarily to be $C^\infty$. (See the example on p. 29 of \cite{carr}.)
It is also in general not unique. However any two centre manifolds
of class $C^k$ are tangent to order $k$ at $x_0$ and in fact the 
coefficients in the Taylor expansion of this submanifold at $x_0$
can be calculated recursively. Examples of this procedure will be 
seen later on.

There is a generalization of the Hartman-Grobman theorem to non-hyperbolic 
stationary points called the reduction theorem \cite{kirchgraber90} which 
reduces the study of the qualitative
behaviour of solutions of a system near a stationary point to
the study of solutions on the centre manifold. In favourable
cases the qualitative behaviour of solutions on the centre
manifold near the origin can be determined once a finite number of
expansion coefficients of the centre manifold are known. This will
be illustrated by the examples of cosmological models treated
later.    

The reduction theorem by itself can provide useful information about
the qualitative behaviour of solutions of a system of ordinary
differential equations. For example this was done for the Mixmaster
solution in \cite{rendall97a}. (Note, however, that since then a much
deeper analysis of this system has been carried out in \cite{ringstrom00a}
and \cite{ringstrom00b}.) This kind of straightforward application of the 
reduction theorem is not what is meant by \lq centre manifold theory\rq\ 
in the present paper. The latter should rather be reserved for the 
procedure where the existence of a centre manifold is used as a basis 
for doing calculations which give concrete information on the nature of
solutions. It is calculations of this type which are the main
tool of this paper.

Centre manifold theory has the advantage of reducing the dimension
of the dynamical system whose qualitative behaviour has to be
determined. The dimension of the centre manifold is of course just
equal to the number of purely imaginary eigenvalues of the 
linearization, counting multiplicity. When the centre manifold is
one-dimensional things become particularly easy. Determining the
stability of the origin in the centre manifold reduces to algebra.
This happens in the example studied in Section \ref{scalar}. When the 
centre manifold is two-dimensional things are a lot more complicated but
there is still a far-reaching theory available (see \cite{perko01},
section 2.11) and in particular cases such as those studied in Sections 
\ref{cmassive} and \ref{cmassless} of this paper, direct approaches may 
lead to the desired result. 

\section{The Bianchi III equations with massive particles}\label{setupmassive}

The results of this section extend the theorems of \cite{rendall00c} and the
fundamental equations will be taken directly from that source. Here it will
be assumed from the start that the Bianchi type is III. The spacetime metric
is of the form
\be\label{eq:metric}
   ds^2 = -dt^2 + a^2(t)(\theta^1)^2 + 
           b^2(t)((\theta^2)^2 + (\theta^3)^2)\ ,
\ee
where the $\theta^i$ are one-forms whose exterior derivatives satisfy the
relations required to ensure that the metric defines a spacetime of Bianchi 
type III. For instance we can assume the relations $d\theta^1=d\theta^2=0$
and $d\theta^3=\theta^2\wedge\theta^3$.
This metric is locally rotationally symmetric (LRS). The matter
content of spacetime is described by the spatially homogeneous phase space
density of particles $f(t,v_i)$. It is assumed that $f$ is invariant under
rotations in the $(v_2,v_3)$-plane and the reflection $v_1\mapsto -v_1$.
The energy-momentum tensor $T_{ij}$ for the Einstein-Vlasov system with 
particles of mass $m\ge 0$ is diagonal and is described by
\bea\label{eq:rho}
  \rho &=& \int f_0(v_i)
  (m^2 + a^{-2}(v_1)^2 + b^{-2}((v_2)^2 + (v_3)^2))^{1/2}
  (ab^2)^{-1} dv_1dv_2dv_3\ ,\nonumber \\
  p_i &=& \int f_0(v_i)g^{ii}(v_i)^2
  (m^2 + a^{-2}(v_1)^2 + b^{-2}((v_2)^2 + (v_3)^2))^{-1/2}
  (ab^2)^{-1} dv_1dv_2dv_3\ ,
\eea 
where $\rho$ is the energy density and $p_i=T^i{}_i$ the pressure components 
of the energy-momentum tensor. The function $f_0$ is determined at some fixed
time $t_0$ by $f_0(v_i)=f(t_0,v_i)$ where $f$ is the phase space density
of particles. Some further technical conditions will be imposed on $f_0$. It 
is assumed to be non-negative and have compact support. It is also assumed 
that the support does not intersect the coordinate planes $v_i=0$. A function 
$f_0$ with this property will be said to have split support. This assumption
ensures that the dynamical system which describes the evolution of the 
spacetimes of interest has smooth coefficients.

The momentum constraints are automatically satisfied for these models.
Only the Hamiltonian constraint and the evolution equations are left.
Instead of considering a set of second order equations in terms of   
$a$ and $b$, 
we will reformulate these equations as a first order system of ODEs by 
introducing a new set of variables. The mean curvature ${\tr} k$
(where $k_{ij}$ is the second fundamental form) is given by
\be
  {\tr} k = -(a^{-1}da/dt + 2b^{-1}db/dt)\ .
\ee
A new dimensionless
time coordinate $\tau$ is defined by $-\fr13\int_{t_0}^t {\tr} k(t)dt$
for some arbitrary fixed time $t_0$. 
In the following a dot over a quantity denotes 
its derivative with respect to $\tau$. 
The Hubble variable $H$ is given by $H = -{\tr} k/3$. This section and the
next are concerned with the case of massive particles ($m>0$). Define the 
following dimensionless variables:
\bea\label{eq:var}
  z &=& m^2/(a^{-2} + 2b^{-2} + m^2)\ ,\nonumber \\
  s &=& b^2/(b^2 + 2a^2)\ ,\nonumber \\
  M_3 &=& 3 b^{-2}({\tr} k)^{-2}\ ,\nonumber \\
  \Sigma_+ &=& -3(b^{-1}db/dt)({\tr} k)^{-1} - 1\ ,
\eea 
These variables lead to a 
decoupling of the equation for the only remaining dimensional variable
$H$ (or equivalently ${\tr} k$)
\be
  \dot{H} = -(1 + q)H\ ,
\ee
where the deceleration parameter $q$ is given by
\be\label{eq:q}
  q = 2\Sigma_+^2 + \fr12 \Omega (1 + R)\ .
\ee
The quantity $R$ is defined by
\be\label{eq:R}
  R = (p_1 + 2p_2)/\rho\ ,
\ee 
where
\bea\label{eq:p}
  p_1/\rho &=& (1 - z)sg_1/h\ ,\nonumber \\
  p_2/\rho &=& \fr12 (1 - z)(1 - s)g_2/h\ ,\nonumber \\
  g_{1,2} &=& \int f_0(v_i) (v_{1,2})^2 
  [z + (1 - z)(s (v_1)^2 + \fr12 (1 - s)((v_2)^2 + (v_3)^2))]^{-1/2}
  dv_1dv_2dv_3\ ,
  \nonumber \\
  h &=& \int f_0(v_i) 
          [z + (1 - z)(s (v_1)^2 + \fr12 (1 - s)((v_2)^2 + (v_3)^2))]^{1/2}
          dv_1dv_2dv_3\ .
\eea

The assumption of split support ensures that the function $R(s,z)$ is a
smooth ($C^\infty$) function of its arguments. The related quantity $R_+$ 
defined by
\be\label{eq:rp}
R_+ = (p_2 - p_1)/\rho
\ee
is a smooth function of $s$ and $z$ for the same reason.

The normalized energy density $\Omega=\rho/(3H^2)$ is determined by 
the Hamiltonian constraint and, in units where $G=1/8\pi$, is given by
\be\label{eq:om}
  \Omega = 1 - \Sigma_+^2 - M_3\ .
\ee 

The assumption of a distribution of massive particles with non-negative 
mass leads to inequalities for $R$, $R_+$ and $\Omega$. Firstly,
$0\le R\le 1$ with $R=0$ only when $z=1$ and $R=1$ only when $z=0$.
Secondly, $-R\le R_+\le \fr12 R$ with $R_+=\fr12 R$ for $s=0$ and $R_+=-R$ 
for $s=1$. Thirdly $\Omega\ge 0$. Using these inequalities in equation 
(\ref{eq:q}) in turn results in $0\leq q \leq 2$ 
(i.e., the same inequality as for causal perfect fluids, see 
\cite{wainwright97}).
 
The remaining dimensionless coupled system is:
\bea\label{eq:eq1}
  \dot{\Sigma}_+ &=& -(2-q)\Sigma_+ + M_3 + \Omega R_+\ ,\nonumber \\
  \dot{s} &=& 6s(1 - s)\Sigma_+\ ,\nonumber \\
  \dot{z} &=& 2z(1 - z)(1 + \Sigma_+ - 3\Sigma_+s)\ ,\nonumber \\
  \dot{M}_3 &=& 2(q - \Sigma_+)M_3\ ,
\eea 
In the following this is referred to for short as the Bianchi III system.
It is of interest to note that the metric functions $a,b$ are expressible 
in terms of $s,z$ in the massive case. The relations are
\be\label{inverse}
  a^2 =  z(m^2s(1-z))^{-1},\quad
  b^2 = 2z(m^2(1-s)(1-z))^{-1}\ .
\ee

There are a number of boundary submanifolds:
\bea\label{eq:sub}
   z &=& 0,1\ ,\nonumber \\
   s &=& 0,1\ ,\nonumber \\    
   \Omega &=& 0\ .
\eea 
The submanifold $z=0$ corresponds to the massless case.
The submanifold $z=1$ leads to a decoupling of the
$s$-equation, leaving a system identical to the corresponding 
dust equations. 
The submanifolds $s = 0,s = 1$ correspond to problems with 
$f_0$ being a distribution while $\Omega = 0$ constitutes the vacuum 
submanifold with test matter.

Including these boundaries yields a compact state space.
In order to apply the standard theory of dynamical systems the coefficients
must be $C^1$ on the entire compact state space $G$ of a given model. 
This is necessary even for uniqueness. In the present case it
suffices to show that $R$ and $R_+$ are $C^1$ on $G$, i.e, 
that they are $C^1$ for $s,z$ when $0\leq s \leq 1\ ,0\leq z \leq 1$. 
As has already been pointed out, this follows from the assumption of
split support, which even implies the analogous statement with $C^1$
replaced by $C^\infty$. It would be possible to get $C^1$ regularity
under the weaker assumption that $f_0$ vanishes as fast as a sufficiently
high power of the distance to the coordinate planes but this is of
little relevance to the main concerns of this paper.

Of key importance is the existence of a monotone function in the 
`massive' interior part of the state space:
\bea\label{eq:mon}
   M &=& (s(1-s)^2)^{-1/3}z(1 - z)^{-1}\ ,\nonumber \\
   \dot{M} &=& 2M\ .
\eea 
Note that the volume $ab^2$ is proportional to $M^{3/2}$. 
This monotone function rules out any interior $\omega$- and $\alpha$-limit
sets and forces these sets to lie on the $s=0$, $s=1$, $z=0$ or $z=1$ parts of 
the boundary.

The physical state space, $G$, of the LRS type III models is given by the 
region in $\R^4$ defined by the inequalities $M_3\ge 0$, 
$0\le s\le 1$, $0\le z\le 1$, and $1 - \Sigma_+^2 - M_3 \geq 0$. A solution
of the Bianchi III system (\ref{eq:eq1}) will be said to lie in the physical
region if it lies in the interior of $G$. In the following we will be 
interested in the behaviour in the limit $\tau\to\infty$ of the solutions of 
the Bianchi III equations which lie in the physical region . This corresponds 
to the asymptotics of the underlying cosmological model in a phase of 
unlimited expansion. It was shown in \cite{rendall00c} that as $\tau$ tends 
to infinity a solution in the physical region converges
to a point of the line of stationary points $L_3$ which consists of all
points of the form $(\fr12,1,z_0,\fr34)$. The point to which the solution
converges must satisfy $z_0>0$. It was left open in \cite{rendall00c} 
whether that point must satisfy $z_0=1$ or whether values less than one
are also possible. In the next section it will be shown that in fact 
$z\to 1$ as $\tau\to\infty$. 

\section{Centre manifold analysis for the Bianchi III equations with massive
particles}\label{cmassive}

It follows from Theorem 5.1 of \cite{rendall00c} that any solution
of the Bianchi III system in the physical region converges to a limit of 
the form $(\fr12,1,z_0,\fr34)$ with $0<z_0\le 1$ as $\tau\to\infty$
and that no solution in the physical region converges to a limit of this 
form as $\tau\to -\infty$. In order to decide which values of $z_0$ can 
occur as limiting values of solutions in the physical region in the forward 
time direction it is necessary to study the behaviour of solutions close to 
the stationary points on the line $L_3$ in detail. If the system is 
linearized about one of these points then it is seen that the linearization 
has two negative eigenvalues and two zero eigenvalues. The fact that there 
is a whole line of stationary points implies that there is automatically 
one zero eigenvalue. However the second zero eigenvalue means that there is 
an extra degeneracy. For $z_0=1$ the matrix defining the linearized 
system is diagonalizable while for $0<z_0<1$ it is not. For this reason 
there are two significantly different cases to be considered.

Consider first the case $z_0=1$. New coordinates will be introduced in order to
simplify the analysis. First the stationary point with $z_0=1$ will be 
translated to the origin. Define $x=\Sigma_+-\fr12$, $y=1-s$,
$w=M_3-\fr34$, $\tilde z=z-1$. Then the transformed system is
\bea
&&\dot x=-\fr14+q(\fr12+x)-2x+w+\Omega R_+          \\
&&\dot y=-6y(1-y)(\fr12+x)                         \\
&&\dot{\tilde z}=-2(1+\tilde z)\tilde z(-2x+\fr32 y+3xy)   \\
&&\dot w=2(q-x-\fr12)(w+\fr34)                     \\ 
&&q=\fr12+2x+2x^2+\fr12\Omega (1+R)                \\
&&\Omega=-x-w-x^2
\eea
Next set $Z=-\tilde z$, $u=x+w$ and $v=x-w$. In order to determine the 
behaviour of 
solutions of the resulting system near the origin we attempt to use the 
linearization at that point. The linearization has zero as an eigenvalue 
with a corresponding two-dimensional eigenspace. This eigenspace is the 
centre subspace. Centre manifold theory \cite{carr} tells us that there is 
an invariant manifold through the origin whose tangent space is the centre 
subspace. This manifold is a centre manifold. A centre manifold does not in 
general need to be unique. However approximations
to it can be derived and these often suffice to determine the qualitative
behaviour of solutions on the centre manifold.
The qualitative behaviour of solutions near the 
stationary point is determined by the qualitative behaviour of solutions on 
any centre manifold. A general property of centre manifolds is that a 
stationary point has an open neighbourhood such that any other stationary 
point in that neighourhood lies on any centre manifold of the original point. 
It follows that in a neighbourhood of the stationary point of the system 
under consideration the part of the line $L_3$ close to that point lies on 
any centre manifold. In the case of present interest the centre manifold can 
be defined by the equations $y=\phi(u,Z)$ and $v=\psi(u,Z)$ where $\phi$ and 
$\psi$ vanish at least quadratically at the origin. The invariance of the 
centre manifold implies that
\bea
&&\dot y=(\d\phi/\d u) \dot u+(\d\phi/\d Z) \dot Z\label{ycentre}      \\
&&\dot v=(\d\psi/\d u) \dot u+(\d\psi/\d Z) \dot Z\label{vcentre}
\eea
on the centre manifold. The right hand sides of these equations vanish to
at least third order at the origin. On the centre manifold
\be
\dot y=-3\phi-3\phi(u+\psi)+3\phi^2(1+u+\psi)\label{ydot}
\ee
Substituting (\ref{ydot}) 
into (\ref{ycentre}) shows that $\phi$ vanishes at quadratic order. 
Putting this information back into the equation (\ref{ycentre}) shows that 
$\phi$ vanishes at third order. This can be repeated indefinitely, with 
the result that $\phi$ vanishes to all orders. On the centre manifold
\bea
&&\dot v=-\fr32\psi-\fr38 u^2+\fr14u\psi+\fr58\psi^2-\fr14 u^3+\fr14 u^2\psi
+u\psi^2+\fr34\psi^3\nonumber       \\
&&+\Omega(R_+-\fr12 R)+\fr14\Omega(1+R)(3\psi-u)
\eea
Note that $(R+R_+)(y,Z)=O(y)$ since $R+R_+$ is smooth and vanishes when
$y=0$. Since $\phi$ vanishes to all orders it follows that $R_+$ and $-R$
are equal to all orders on the centre manifold. Hence to quadratic order 
$\psi(u,Z)=-\fr1{12}u^2-r_2 uZ+\ldots$ where $r_2$ is the derivative of $R$ 
with respect to $z$, evaluated at the stationary point of interest. Next 
the dynamics on the centre manifold will be examined. 
\bea
&&\dot u=u^2+\fr{11}4 u\psi+\fr34\psi^2+\fr9{16}u^3+\fr{15}{16}u^2\psi
+\fr3{16}u\psi^2-\fr3{16}\psi^3\nonumber         \\
&&+\Omega[(R+R_+)+\fr14 R(3u-\psi)]
\eea
while
\be
\dot Z=2uZ-2uZ^2+2Z\psi-2Z^2\psi-3Z(1-Z)\phi(1+u+\psi)
\ee
Discarding terms of order greater than two gives the truncated system
\bea
&\dot u=u^2    \\
&\dot Z=2uZ
\eea
For this system $u^{-2}Z$ is a conserved quantity and so the qualitative
behaviour of the solutions is easily determined. Is the qualitative 
behaviour of solutions of the full system on the centre manifold similar? 
The terms on the right hand side of the restriction of the evolution
equations to the centre manifold written out explicitly above contain a 
factor of $u$. This is not an accident and in fact the full 
equations are of the form
\bea
&&\dot u=uf(u,Z)     \\
&&\dot Z=ug(u,Z)
\eea 
for some functions $f$ and $g$ of any arbitrary finite degree of 
differentiability. This is because of the fact, already
mentioned above, that the line $L_3$ of stationary points is locally 
contained in the centre manifold. Thus the restriction of the system to the 
centre manifold has a line of stationary points with $u=0$. This means that 
the vector field defining the dynamical system vanishes for $u=0$. The 
existence of the functions $f$ and $g$ then follows from Taylor's theorem. 
The fact that $L_3$ lies on the centre manifold also implies that 
$\psi(u,Z)=uh(u,Z)$ for a function $h$ of arbitrary finite 
differentiability.

It is possible to get more precise information about the function $f$ as
follows. Recall that $R+R_+$ vanishes modulo the function $\phi$. Since 
$\phi$ vanishes to all orders it follows that expressions containing a 
factor $R+R_+$ can be ignored in all perturbative calculations. We have
\be
\Omega=-u-\fr14 u^2-\fr12 u\psi-\fr14\psi^2
\ee
It follows that every term in the evolution equation for $u$ on the centre
manifold contains a factor of $u^2$, either directly or via the fact that
$\psi=uh$. Hence $f(0,Z)=0$. 

For the purpose of qualitative analysis we can replace the above system 
by the system
\bea
&&u'=f(u,Z)+\ldots     \\
&&Z'=g(u,Z)+\ldots
\eea 
which has the same integral curves away from the $Z$-axis. Only solutions 
with $Z\ge 0$ are physical. The $Z$-axis is an invariant manifold of the 
rescaled dynamical system. This new system has a hyperbolic stationary 
point at the origin which is a source. This makes it easy to determine the 
phase portrait for the system 
near the endpoint of $L_3$ at $z=1$. What we see is that there is an open set 
of initial data in the physical region such that the corresponding solutions 
approach the stationary point with $z=1$ as $\tau\to\infty$ and there is a 
neighbourhood of the endpoint of $L_3$ at $z=1$ such that in that 
neighbourhood no solution in the physical region approaches a point of 
$L_3$ with $z<1$. 

Consider now the case $0<z_0<1$. The analysis is similar to that in the
case already considered, although the calculations are somewhat heavier.
For $z_0<1$ the linearization has a 
non-diagonal Jordan form.  Define $x=\Sigma_+-\fr12$, $y=1-s$,
$w=M_3-\fr34$, $\tilde z=z-z_0$. Then the transformed system is
\bea\label{transformed}
&&\dot x=-\fr14+q(\fr12+x)-2x+w+\Omega R_+          \\
&&\dot y=-6y(1-y)(\fr12+x)                         \\
&&\dot{\tilde z}=2(z_0+\tilde z)(1-z_0-\tilde z)(-2x+\fr32 y+3xy)   \\
&&\dot w=2(q-x-\fr12)(w+\fr34)                     \\ 
&&q=\fr12+2x+2x^2+\fr12\Omega (1+R)                \\
&&\Omega=-x-w-x^2
\eea
The linearization has zero as an eigenvalue with a corresponding 
two-dimensional generalized eigenspace. 

For calculational purposes it is convenient to choose coordinates which reduce
the linearization to a simple form. For this reason the following variables
will be defined. They almost reduce the linearized operator to Jordan form.
\bea
&&U=-2z_0(1-z_0)(x+w)                              \\
&&V=-\fr12 [(\bar R-1)x+(\bar R+1)w]               \\
&&Y=y                                              \\
&&Z=\tilde z +z_0(1-z_0)y+\fr43 [(\bar R-1)x+(\bar R+1)w]z_0(1-z_0)
\eea  
Here $\bar R$ is the value of $R$ when $s=1$ and $z=z_0$. Note that the
value of $R_+$ when $s=1$ and $z=z_0$ is $-\bar R$. Up to linear order
the equations in the new coordinates are
\bea\label{linear3}
&&\dot U=0+\ldots                    \\
&&\dot V=-\fr32 V+\ldots             \\
&&\dot Y=-3Y+\ldots                  \\
&&\dot Z=(\bar R+1)U+\ldots
\eea
 
In the new coordinates the centre subspace is defined by $V=0$ and $Y=0$.
The centre manifold is defined by equations of the form $Y=\phi(U,Z)$ and
$V=\psi(U,Z)$, where $\phi$ and $\psi$ are functions of any desired finite 
degree of differentiability which vanish at least quadratically at the 
origin. (These are not the same functions $\phi$ and $\psi$ introduced
in the centre manifold calculation for $z_0=1$.) Now apply the equations
\bea
&&\dot Y=(\d\phi/\d U) \dot U+(\d\phi/\d Z) \dot Z      \\
&&\dot V=(\d\psi/\d U) \dot U+(\d\psi/\d Z) \dot Z
\eea
which must hold on the centre manifold. The linear contribution to
$\dot Z$ is proportional to $U$ and hence the quadratic contribution
to the right hand side of the first equation contains only terms 
proportional to $U^2$ and $UZ$ and no term proportional to $Z^2$.
The quadratic part of the left hand side is proportional to the
quadratic part of $\phi$. Suppose that the latter is given by 
$AU^2+BUZ+CZ^2$ for constants $A$, $B$ and $C$. Comparing the
coefficients of $Z^2$ shows that $C=0$. Then comparing coefficients
of $UZ$ shows that $B=0$. At this point it has been shown that the
right hand side of the equation vanishes to quadratic order and it
follows that $A=0$. Hence $\phi$ vanishes to quadratic order. As in
the case $z_0=1$ this information can be used repeatedly to show that
$\phi$ vanishes to all orders at the origin.

Next the equation for $\dot V$ on the centre manifold will be examined. 
It turns out that the quadratic contribution to $\dot V$ is a linear 
combination of $U^2$ and $UZ$. It can be concluded that up to
quadratic order $\psi=\alpha U^2+\beta UZ+\ldots$ for some constants 
$\alpha$ and $\beta$. These 
constants depend on the value of $z_0$ at the point where the linearization 
is carried out and on the initial data $f_0$. 

Substituting the 
information about the centre manifold into the original system shows 
that the restriction of the dynamical system to the centre manifold 
has the following form up to quadratic order:
\bea\label{quadratic3}
&&\dot U=\gamma U^2+ \ldots     \\
&&\dot Z=(\bar R+1)U+\delta U^2+\epsilon UZ+\ldots
\eea
for some constants $\gamma$, $\delta$ and $\epsilon$.

All the terms on the right hand side of the restriction of the evolution
equations to the centre manifold written out explicitly above contain a 
factor of $U$ and for the same reason as in the case $z_0=1$
the full equations are of the form
\bea
&&\dot U=Uf(U,Z)     \\
&&\dot Z=Ug(U,Z)
\eea 
for some functions $f$ and $g$ of arbitrary finite differentiability. Thus 
the restriction of the system to the centre manifold has a line of stationary 
points with $U=0$. 

Consider now the dynamical system defined by
\bea
&&U'=f(U,Z)     \\
&&Z'=g(U,Z)
\eea 
where the prime denotes $d/d\sigma$ for a time coordinate $\sigma$.
This has the same integral curves as the system we want to analyse
in the complement of the $Z$-axis. It will be referred to in the following
as the rescaled system. At the origin $f$ vanishes while
$g$ has the non-vanishing value $\bar R+1$. In fact $f(0,Z)=0$ for all
$Z$, the argument being just as in the case $z_0=1$. As a consequence 
the integral curve of the rescaled system passing through the origin
lies on the $Z$-axis. Since the vector field defining this system is 
non-vanishing at the origin it is clear that no solution can approach 
the origin from the physical region.

Putting together the results of the two centre manifold analyses with
Theorem 5.1 of \cite{rendall00c} yields
the following theorem:

\noindent
{\bf Theorem 1} If a smooth non-vacuum reflection-symmetric LRS solution of 
Bianchi type III of the Einstein-Vlasov equations for massive particles is 
represented as a solution of (\ref{eq:eq1}) then for $\tau\to\infty$ it
converges to the point with coordinates $(\fr12,1,1,\fr34)$. The quantities
$p_1/\rho$ and $p_2/\rho$ converge to zero as $t\to\infty$.

Next some information will be obtained on the detailed asymptotic behaviour
of the spacetime as $t\to\infty$. Consider first the behaviour of a 
solution of the rescaled dynamical system for $z_0=1$ as $\sigma\to -\infty$. 
At the origin the rescaled system has a stationary point which is a  
hyperbolic source. By a theorem of Hartman (see \cite{perko01}, p. 127)
the system can be linearized by a $C^1$ mapping. As a consequence
$u=u_0 e^{\sigma}+o(e^{\sigma})$ for some constant $u_0$ and 
$Z=o(e^{\sigma})$. Along a solution of the equations 
$d\tau/d\sigma=u^{-1}(\sigma)$. Hence for large negative values of $\sigma$  
we have $\tau=-u_0^{-1}e^{-\sigma}+o(e^{-\sigma})$. Hence on the centre 
manifold $u=-\tau^{-1}+o(\tau^{-1})$ and $Z=o(\tau^{-1})$.

General results on centre manifolds \cite{carr} show that the general solution
satisfies
\bea
&&u(\tau)=-\tau^{-1}+o(\tau^{-1})     \\
&&z(\tau)=1+o(\tau^{-1})                \\
&&y(\tau)=o(\tau^{-k})                  \\
&&v(\tau)=-\fr1{12}\tau^{-2}+o(\tau^{-2})
\eea
where $k$ is any positive integer. As
$\tau\to\infty$ we have $q=\fr12+o(1)$. Now $dH/dt=-(1+q)H^2$. 
It follows that $H=\fr23 t^{-1}(1+o(1))$. Thus in leading order 
$\tau=\fr23\log t+\dots$. Putting this information in the
asymptotic expressions above gives the leading order behaviour of
$u$, $z$, $y$ and $v$ as functions of $t$. This in turn gives the
asymptotic behaviour of the variables $\Sigma_+$, $s$, $M_3$ and
$z$. The result is
\bea
&\Sigma_+(t)=\fr12-\fr34 (\log t)^{-1}+\ldots   \\
&s=1+o((\log t)^{-k})                                  \\
&M_3=\fr34-\fr34 (\log t)^{-1}+\ldots          \\
&z=1+o((\log t)^{-1})
\eea
From this it can be concluded using (\ref{inverse}) that 
$a/(\log t)^{1/2}$ goes to infinity
as $t\to\infty$. It follows directly from the fundamental
equations that $a$ goes to infinity slower than any power of $t$.
To see this note that since $\Sigma_+$ tends to $\fr12$, the
quantity $(b^{-1}db/dt)(\tr k)^{-1}$ tends to $-\fr12$. Hence
$b^{-1}db/dt=t^{-1}+\ldots$ and $a^{-1}da/dt=o(t^{-1})$.
This gives the desired conclusion concerning the upper bound for
the rate of growth of $a$. The leading order behaviour of $b$ can be read 
off from the defining equation for $M_3$, with the result that 
$b=t+\ldots$.

\section{Centre manifold analysis for the Bianchi III equations with massless
particles}\label{cmassless}

The analysis of LRS Bianchi spacetimes with massless particles in 
\cite{rendall99a} included the case of Bianchi type III. In this section 
the results obtained in \cite{rendall99a} will be sharpened. As previously 
indicated, the dynamical system  for LRS Bianchi 
type III spacetimes with massless particles is equivalent to the boundary
component $z=0$ of the region $G$ on which the system for massive particles
is defined. Here we adopt the notation of \cite{rendall00c} rather than 
that of \cite{rendall99a}. Then the dynamical system required in this section
can be obtained by specializing a system we had in the last section. The 
result is
\bea
&&\dot x=-\fr14+q(\fr12+x)-2x+w+\Omega R_+          \\
&&\dot y=-6y(1-y)(\fr12+x)                          \\
&&\dot w=2(q-x-\fr12)(w+\fr34)                      \\ 
&&q=\fr12+2x+2x^2+\Omega                            \\
&&\Omega=-x-w-x^2
\eea 
The only modifications are that the equation for $z$ has been dropped and
that the relation $R=1$ has been used. This is a three-dimensional dynamical
system. Its linearization at the origin has eigenvalues $0$, $-\fr32$ and 
$-3$ and its kernel is spanned by $(1,0,0)$. It follows that this point has 
a one-dimensional centre manifold. It was already shown in \cite{rendall99a} 
that all solutions of this system corresponding to solutions of the 
Einstein-Vlasov equations with massless particles converge to the origin 
as $\tau\to\infty$. The purpose of the centre manifold analysis here is to 
obtain more detailed information about how this point is approached and hence 
how the scale factors behave in the phase of unlimited expansion. It is useful
to substitute in the expressions for $\Omega$ and $q$, thus obtaining the
following dynamical system:
\bea
&&\dot x=\fr32 w+\fr52 x^2-xw+x^3-(x+w+x^2)(1+R_+)  \\
&&\dot y=-6y(1-y)(\fr12+x)                          \\
&&\dot w=2(-w+x^2)(w+\fr34)
\eea
The centre manifold is defined by the equations $y=\phi(x)$ and $w=\psi(x)$. 
Using the same procedure as in the last section the functions $\phi$ and 
$\psi$ can be determined up to quadratic order. In this case it turns out
that $\phi$ vanishes to quadratic order while to that order $\psi=x^2+\ldots$.
Substituting this information into the evolution equation for $x$ shows 
that $\dot x=4x^2+O(|x|^3)$ on the centre manifold. Since physical solutions
converge to the origin as $\tau\to\infty$ they must be approximated by 
solutions on the centre manifold with $x$ negative. It can be concluded 
that for these solutions $\Sigma_+$ is eventually less than one half.

With the information already obtained, various features of the dynamics
of the spacetime can be reconstructed. To leading order
$x=-(1/4\tau)+\ldots$, $y=o(\tau^{-2})$ and $w=1/16\tau^2+o(\tau^{-2})$. 
The quantity $q$ tends to $\fr12$ and so some of the calculations for the
massive case can be taken over here. As in that case $b=t+\ldots$. It is
not possible to determine the asymptotic behaviour of the scale factor 
$a$ in the same way as in the massive case; a different approach is 
necessary. It may be computed that $a^{-1}da/dt=\fr23(\tr k)x$. Hence
in leading order $a^{-1}da/dt=\fr12 (t\log t)^{-1}+\ldots$. It follows 
that $a\to\infty$ as $t\to\infty$.

\section{Inflation with a massive scalar field}\label{scalar}

This section presents a rigorous derivation of some heuristic results 
of \cite{belinskii86} for spatially flat homogeneous and
isotropic spacetimes with a massive scalar field as source. The
starting point is the following system of equations from \cite{belinskii86}.
\bea
&x_\eta=y                           \\
&y_\eta=-x-3y(x^2+y^2)^{1/2}
\eea
This is a dynamical system on the plane whose coefficients are smooth
everywhere except at the origin, where they are $C^1$. Transforming 
to polar coordinates $(r,\theta)$ and introducing $\rho=r/(1+r)$ gives 
the system
\bea
&\rho_\eta=-3\rho^2\sin^2\theta     \\
&\theta_\eta=-1-3\rho(1-\rho)^{-1}\sin\theta\cos\theta
\eea
Introducing a new time coordinate $\tau$ and setting $u=1-\rho$ gives
\bea
&u_\tau=3u(1-u)^2\sin^2\theta        \\
&\theta_\tau=-u-3(1-u)\sin\theta\cos\theta
\eea
So far this is nothing new compared to what is done in \cite{belinskii86}.
Now the stationary point at the origin will be investigated. The 
eigenvalues of the linearization are $-3$ and zero. Evidently the 
$\theta$-axis is invariant and in fact it is the stable manifold of
the origin. The centre subspace is given by $u+3\theta=0$ and thus we
introduce $v=u+3\theta$ in order to study the centre manifold. This leads to
the transformed system
\bea
&u_\tau=3u(1-u)^2\sin^2(\fr13(v-u))     \\
&v_\tau=-3u\cos^2(\fr13(v-u))-6u^2\sin^2(\fr13(v-u))
+3u^3\sin^2(\fr13(v-u))\nonumber                 \\
&-9\sin(\fr13(v-u))\cos(\fr13(v-u))+9u\sin(\fr13(v-u))\cos(\fr13(v-u))
\eea
The centre manifold is of the form $v=\phi(u)$. To quadratic order
$\phi(u)=-u^2+\ldots$. On the centre manifold $u_\tau=\fr13 u^3+O(u^4)$ 
and hence there is a unique solution which enters the physical region.

The asymptotic form of the solution as $\tau\to -\infty$ is 
$u=\sqrt{3/2}(-\tau)^{-1/2}+\ldots$. It follows that $v=\fr32\tau^{-1}+\ldots$.
Hence
\be
\rho=1-\sqrt{3/2}(-\tau)^{-1/2}+\ldots
\ee
and 
\be
\theta=-\sqrt{1/6}(-\tau)^{-1/2}+\ldots
\ee
The next step is to convert back to the original variables. In leading
order $\eta=-\sqrt{6}(-\tau)^{1/2}+\ldots$ and so $\rho=1+3/\eta+\ldots$
and $\theta=1/6\eta+\ldots$. It follows that $x=\eta/3+\ldots$ and
$y=\fr1{18}+\ldots$. The combination $z=\sqrt{x^2+y^2}$ occurring in
\cite{belinskii86} is asymptotic to $\eta/3$. The linear dependence
of $x$ and $z$ with respect to $\eta$ is what was found in 
\cite{belinskii86}. The interpretation of these variables is
that $\eta$, $x$ and $z$ are proportional to proper time, the scalar 
field $\phi$ and the mean  curvature of the hypersurfaces of constant 
time respectively. Thus in the limit $t\to -\infty$ both $\phi$ and
the mean curvature are proportional to $t$. The leading order 
behaviour of the scale factor for large negative times is, up to
inessential constants, $e^{t^2}$. This gives a rigorous confirmation
of the conclusions of \cite{belinskii86}. Note that the relevance of
centre manifolds in this context has been pointed out by Foster
\cite{foster98} but that he did not carry out a full centre manifold
analysis; his paper was mainly concerned with other applications of
dynamical systems to homogeneous and isotropic solutions of the Einstein
equations coupled to a scalar field with potential.

\section{Further comments}

This section contains some further comments on the results of Sections 
\ref{cmassive} and \ref{cmassless}. In terms of the dimensionless
variables which are the unknowns in the dynamical system studied
in those sections, solutions of the equations with collisionless
matter (with both massless and massive particles) converge at late
times to a point corresponding to a vacuum solution of Bianchi type 
III. The corresponding statement holds for LRS Bianchi type III
solutions with dust. (Cf. the discussion in section 6 of
\cite{rendall99a}.) In particular, the dimensionless density
parameter $\Omega$ tends to zero in the expanding direction.
Furthermore, in the case of massive particles, the ratios $T^i_i/\rho$ 
of the spatial eigenvalues of
the energy-momentum tensor (i.e. the principal pressures) to the
energy density tend to zero. Thus in a certain sense both solutions
with collisionless matter and dust solutions are approximated by
vacuum solutions, while solutions with collisionless matter and
massive particles are approximated better by dust solutions than 
either of these are approximated by vacuum solutions. This is one
of the main results of this paper. 
 
One of the scale factors, $a$, grows much more slowly than the other
scale factor $b$. An important conclusion is that $a$ tends to
infinity at all in the case of collisionless matter, both in the 
massive and massless cases. Centre manifold theory was the essential 
tool which allowed us to prove this and thus to go beyond what was
achieved in \cite{rendall99a} and \cite{rendall00c}. In this sense 
both types of collisionless matter resemble each other and differ
from the vacuum case, where $a$ is asymptotically constant. These
results lend further support to the following two suggestions made 
in \cite{rendall99a} and \cite{rendall00c}. The first is that any
spatially homogeneous solution of the Einstein equations with
collisionless matter and massive particles behaves asymptotically
in a phase of unlimited expansion like a dust model. The second is
that in any non-vacuum spatially homogeneous model with perfect
fluid or collisionless matter eventually every scale factor
increases in the time direction in which the volume increases. 
It would be interesting to investigate the truth of these statements 
in solutions of the Einstein equations with collisionless matter
which are LRS Bianchi type VIII or type II but not LRS.

\end{document}